# 3D logic cells design and results based on Vertical NWFET technology including tied compact model

Chhandak Mukherjee, Marina Deng, François Marc, Cristell Maneux
IMS Laboratory
University of Bordeaux, CNRS UMR 5218, Bordeaux INP
Talence, France
firstname.name@u-bordeaux.fr

Arnaud Poittevin, Ian O'Connor, Sébastien Le Beux
Lyon Institute of Nanotechnology
University of Lyon, CNRS UMR 5270, Ecole Centrale de Lyon
Ecully, France
firstname.name@ec-lyon.fr

Abhishek Kumar, Aurélie Lecestre, Guilhem Larrieu
LAAS-CNRS
Université de Toulouse, CNRS, INP
Toulouse, France
firstname.name@laas.fr

*Abstract*— **Gate-all-around Vertical Nanowire Field Effect Transistors (VNWFET) are emerging devices, which are well suited to pursue scaling beyond lateral scaling limitations around 7nm. This work explores the relative merits and drawbacks of the technology in the context of logic cell design. We describe a junctionless nanowire technology and associated compact model, which accurately describes fabricated device behavior in all regions of operations for transistors based on between 16 and 625 parallel nanowires of diameters between 22 and 50nm. We used this model to simulate the projected performance of inverter logic gates based on passive load, active load and complementary topologies and carry out an performance exploration for the number of nanowires in transistors. In terms of compactness, through a dedicated full 3D layout design, we also demonstrate a 1.4x reduction in lateral dimensions for the complementary structure with respect to 7nm FinFET-based inverters.**

***Keywords—Vertical NWFET technology, compact model, VNWFET DC measurements, 3D logic circuit cell, circuit simulation results.***

## I. INTRODUCTION

Data size and functionality requirements for computing are increasing, according to the expectation that hardware performance will continue to improve, irrespective of the actual implementation. This is particularly true for emerging computing paradigms such as Edge Computing which is placing extraordinarily stringent constraints on computing hardware performances. However, the end of the roadmapped technological scaling is anticipated in a few technology nodes, mainly for cost reasons down to the 7nm FinFET gate length node. In this context, vertical integration is an attractive approach to fully take advantage of 3D integration and scale pitch between contacts. Huge gains in silicon area are expected through the combination of extremely small elementary device footprint and minimal device usage with MIG and PTL design styles, for instance. This paper is the first attempt to quantify the gains in terms of compactness and energy efficiency of 3D logic blocks based on actual fabricated p-type VNWFET devices.

The paper is organized as follows: section II recalls the VNWFET technology main features while detailing its associated scalable compact model. In particular, the unified charge-based control model has been self-consistently modified to take into account depletion and accumulation regimes, electrostatic control, short-channel effects (SCE), drain-induced barrier lowering (DIBL) and band-to-band tunneling (BTBT) contributions through gate-induced drain leakage (GIBL). Simulated results are compared to measurements to illustrate the p-type VNWFET model versatility in terms of dimensions. An n-type VNWFET model has also been delivered using the electron mobility value from the literature. These scalable compact models have been implemented in Verilog-A, and subsequently implemented in a dedicated circuit design workspace. In section III, we demonstrate the efficiency of this design workspace to simulate and quantify the 3D logic blocks. 3D layouts implement inverter functions with various topologies: (i) passive load, (ii) active load and (iii) complementary. Their static and dynamic energy consumption and delays are given. In section IV, we propose a layout footprint comparison between the 7nm FinFET and the VNWFET through conventional λ-rules. Going beyond this approach, section V deals with large-scale integration considerations suitable for a fully 3D logic block architecture.

## II. VNWFET DEVICES

### A. Technology description

The VNWFET technology has a junction-less architecture composed of a homogenous highly doped nanowire channel, patterned into boron doped ($2\times10^{19}cm^{-3}$) Si substrate. The current flows between silicided source/drain contacts and is controlled by a gate-all-around structure with a physical channel length of 14nm (fig. 1). More details on the fabrication steps can be found in [2].

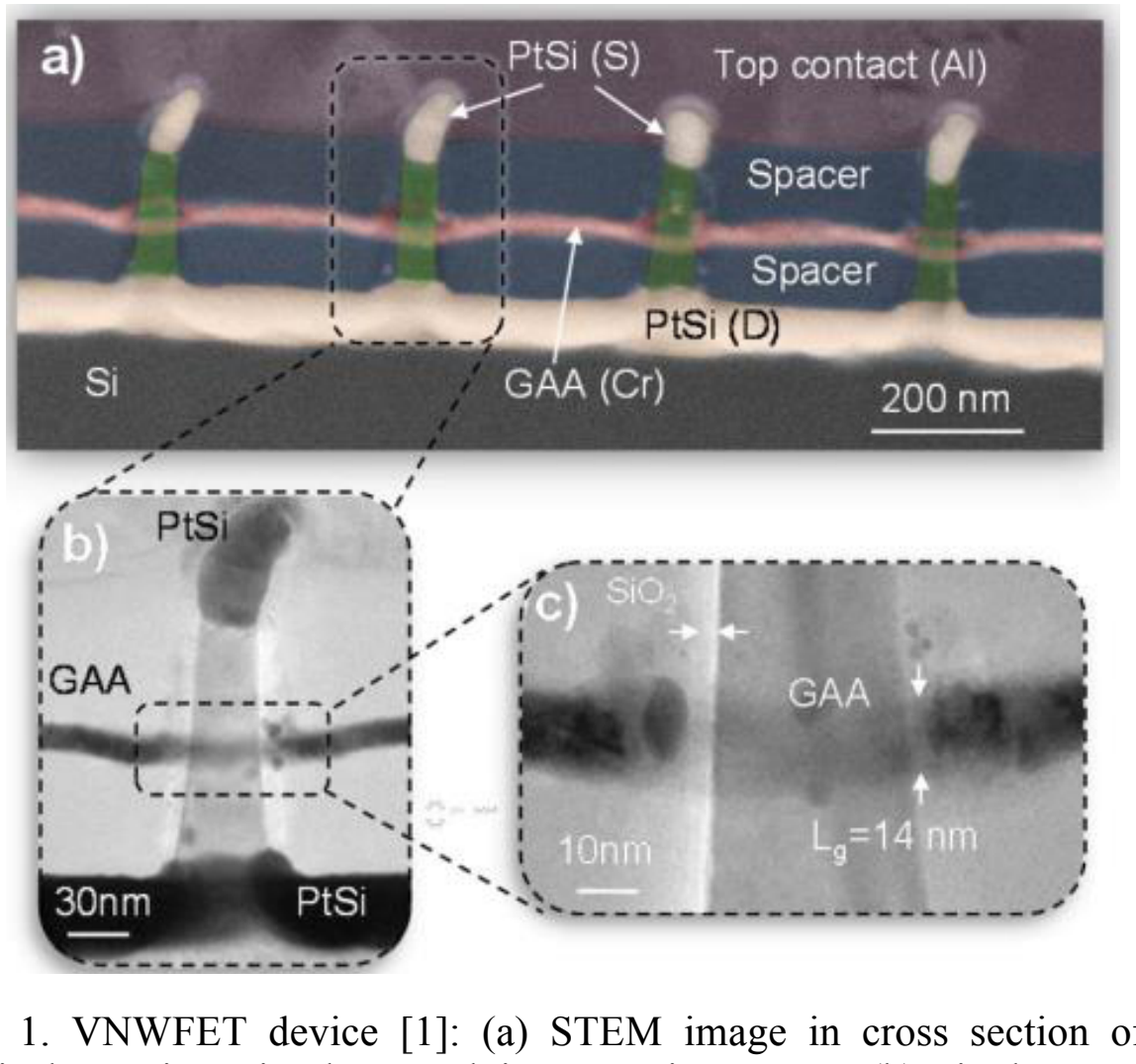

Fig. 1. VNWFET device [1]: (a) STEM image in cross section of the vertical transistor implemented in nanowire arrays, (b) single nanowire showing its (c) gate formation.

## B. Compact Model

The model formulation is based on the unified charge-based control model (UCCM) elaborated in [3] for long-channel devices, which furthers the physical basis of the junctionless nanowire transistor (JLNT) model presented in [4] and adapted in [5] for the JLNT technology under test [2]. The limitations of the model in [4] is mainly the piece-wise continuous drain-current model which requires additional smoothing functions and fitting parameters to smooth the transition between depletion and accumulation modes of operation. In order to overcome this, the explicit and non-piece-wise solution in [3] treats the mobile charge ($Q_m$) to be decoupled between the depletion ($Q_{DP}$) and complementary ($Q_C$) components. In the depletion mode the UCCM expression has been formulated as [3]:

$$Q_{DP} = Q_{eff} LW \left\{ \frac{Q_{sc}}{Q_{eff}} \exp\left( \frac{V_g - V_{th} - \eta V}{\eta \phi_T} + \frac{Q_{dep}}{Q_{sc}} \right) \right\} \qquad (1)$$

with the depletion charge, $Q_{dep}=qN_DR/2$, the effective charge during depletion, $Q_{eff}=Q_{sc}\eta C_{ox}\varphi_T/(Q_{sc}+\eta C_{ox}\varphi_T)$, $Q_{sc}=2\varepsilon_{Si}\varphi_T/R$, $R$ being the nanowire diameter, $\eta$ an interface trap parameter, $\varphi_T$ the thermal voltage and $V$ the potential along the channel. A Lambert W function has been used in both [3] and [4] to develop the solution of total mobile charge in the JLNT. While the expression for $Q_{DP}$ predicts the depletion contribution correctly (for $V_g<V_{th}$), it underestimates the value of the drain current above the flat-band condition. So in accumulation mode, especially in high accumulation with $Q_C \geq Q_{dep}$, the charge $Q_C$ has been derived to act complementary to $Q_{DP}$, considering that the threshold voltage is pinned at $V_{FB}$ in the accumulation region, in order to avoid using additional smoothing functions and improve simulation time. Under high accumulation $Q_C \geq Q_{dep}$ and $Q_C$ is simplified using another Lambert function as follows [3]:

$$Q_C = \eta C_c \phi_T LW \left\{ \frac{Q_{sc}}{\eta C_c \phi_T} \exp\left( \frac{V_g - V_{FB} - \eta V}{\eta \phi_T} \right) \right\} \qquad (2)$$

with corrected electrostatic control through $C_c=C_{ox}-C_{eff}$, $C_{eff}=1/C_{ox}+R/2\varepsilon_{Si}$. Having evaluated both the depletion and complementary parts of the mobile charge, one can formulate the non-piece-wise continuous model of the total drain current in terms of $Q_{DP}$ and $Q_{DC}$ at the source and the drain end, $Q_{DP0}$, $Q_{C0}$ and $Q_{DPL}$, $Q_{CL}$, respectively:

$$I_{DS,0} = \mu_{eff} \frac{2\pi R}{L_{eff}} \phi_T \left[ \frac{Q_{DP}^2}{2\eta C_{ox}\phi_T} + Q_{DP} + \frac{Q_C^2}{2\eta C_c \phi_T} + 2Q_C \right]_{Q_{DPL},\, Q_{CL}}^{Q_{DP0},\, Q_{C0}} \qquad (3)$$

The drain current expression is free of any fitting parameters and can be evaluated based on the physical device parameters such as that of geometry and doping. Additionally, short channel effects were taken into account considering velocity saturation, an effective mobility, $\mu_{eff}$, and incorporating an effective gate length, $L_{eff}=L-\Delta L$, where $L$ is the physical device gate length and $\Delta L$ is calculated following the expression in [6]. Considering that the source and drain access region resistances degrade the drain current above threshold, the final expression of the drain current can be written as a function of the long channel current ($I_{DS,0}$), using (3), taking into account the corrections due to short-channel effects [5], as follows [6]:

$$I_{DS} = \frac{I_{DS,0} NF}{1 + 2\pi \frac{R}{L_{eff}} NF \mu_{eff} (R_S + R_D) \left[ (Q_{DP0} + Q_{C0}) - \eta_1 \left( Q_{DP0} + Q_{C0} - (Q_{DP,VDeff} + Q_{C,VDeff}) \right) \right]} \qquad (4)$$

Here, $R_S$ and $R_D$ are the source and drain series access resistances, respectively; $NF$ is the number of nanowires in parallel, $\eta_1$ is a fine tuning parameter to take into account the drain-voltage dependence of the series access resistances and $Q_{DP,Vdeff}+Q_{C,Vdeff}$ is the total mobile charge at the drain end (pinch-off) of the channel.

Additionally, considering formation of Schottky contacts at the source and drain access regions, the subthreshold leakage currents are also taken into account. Consequently, thermionic ($I_{th}$), tunneling ($I_{tun}$) and band-to-band tunneling (BTBT) contributions through gate-induced drain leakage (GIDL) are added as separate branch currents [7] to the total drain current, in order to model the subthreshold behavior of the drain current. The expression used in the compact model for the BTBT current at the drain end reads [7]:

$$I_{GIDL} = 2\pi R L_{Access} NF \cdot A_{GIDL} V_{DS} E_{segd}^{\,2} \exp\left( -\frac{B_{GIDL}}{E_{segd}} \right) \qquad (5)$$

wher $L_{Access}$ represents the lengths of the source and drain access regions outside the channel, $B_{GIDL}$ is a physics-based parameter with a theoretical value of 21.3MV/cm [7] and $E_{segd}$ is the electric field in the drain overlap region, given as

$$E_{segd} = \frac{C_{ox}\sqrt{V_{segd}^{\,2} + (C_{GIDL} V_{DS})^2}}{\varepsilon_0 \varepsilon_{Si}} \qquad (6)$$

Here, $V_{segd}$ is the gate-drain voltage across the oxide and $A_{GIDL}$, $C_{GIDL}$ are two GIDL fitting parameters. Lastly, additional model improvement has been achieved compared to the model reported in [5], in the subthreshold regime. In order to improve model accuracy, the accurate extraction of the parameter $\eta$ is ensured in order to correctly adjust the subthreshold slope. Moreover, the effect of drain-induced barrier lowering (DIBL) is also taken into account in the compact model by a modification of the threshold voltage through the following equation,

$$V_{th} = V_{FB} - \frac{Q_{dep}}{C_{ox}} - DIBL(V_{DSmax} - V_{DSmin}) \qquad (7)$$

with $DIBL$ being the drain-induced barrier lowering in mV/V.

## C. Measured and simulated results

For the validation of the compact model against measurement results, we chose a wide range of geometries where test structures had diameters ($D$) ranging between 22-50nm with 16-625 nanowires in parallel ($NF$). Figs. 2(a) and (b) show the transfer characteristics, $I_D$-$V_{GS}$, of the JLNTs with D=22nm/NF=16 and D=50nm/NF=36, respectively. The model simulation results show very good agreement with the measurements over the entire bias range, indicating accuracy of individual modules of the compact model. Particularly, the improvement of the model accuracy in the subthreshold region is observable compared to the results reported in [5], leveraging eqs. (5) and (7) as well as the parameter $\eta$. The improvement in drive current and subthreshold leakage with a higher number of nanowires in parallel is obvious from fig. 2(b), which however suffers from a more pronounced DIBL induced $V_T$-shift. This is most likely due to quantum confinement effects in smaller nanowire diameters [8]. Nevertheless, the compact model captures these effects with sufficient accuracy. A second order validation is performed in figs. 3 (a) and (b) depicting the output characteristics, $I_D$-$V_{DS}$, of the JLNTs, further affirming model accuracy.

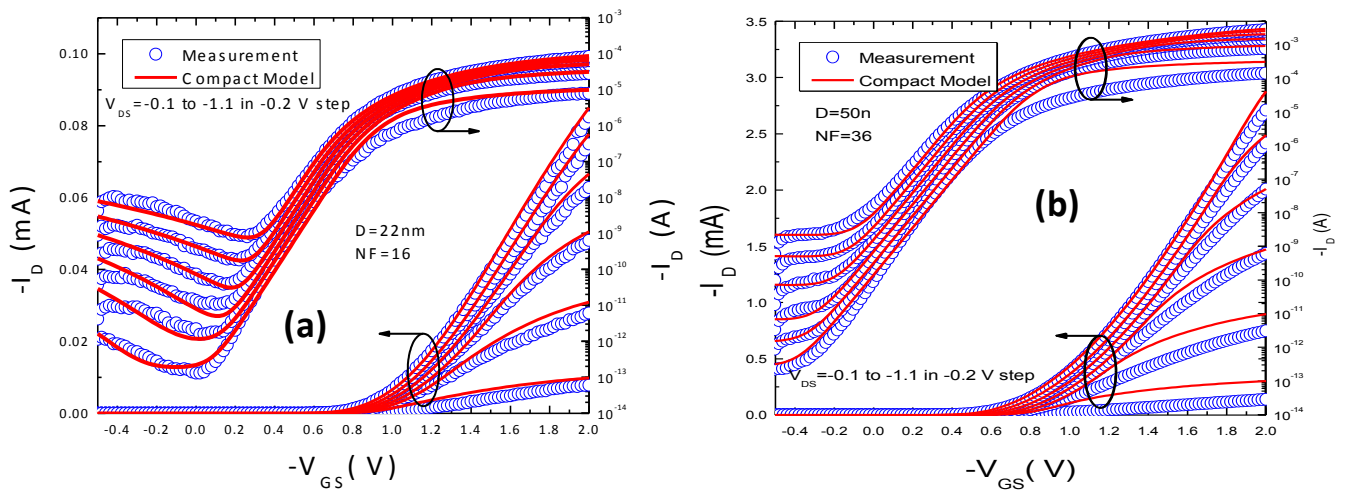

Fig. 2: $I_D$-$V_{GS}$ of JLNTs with (a) 22nm diameter and 16 nanowires in parallel and (b) 50nm diameter and 36 nanowires in parallel.

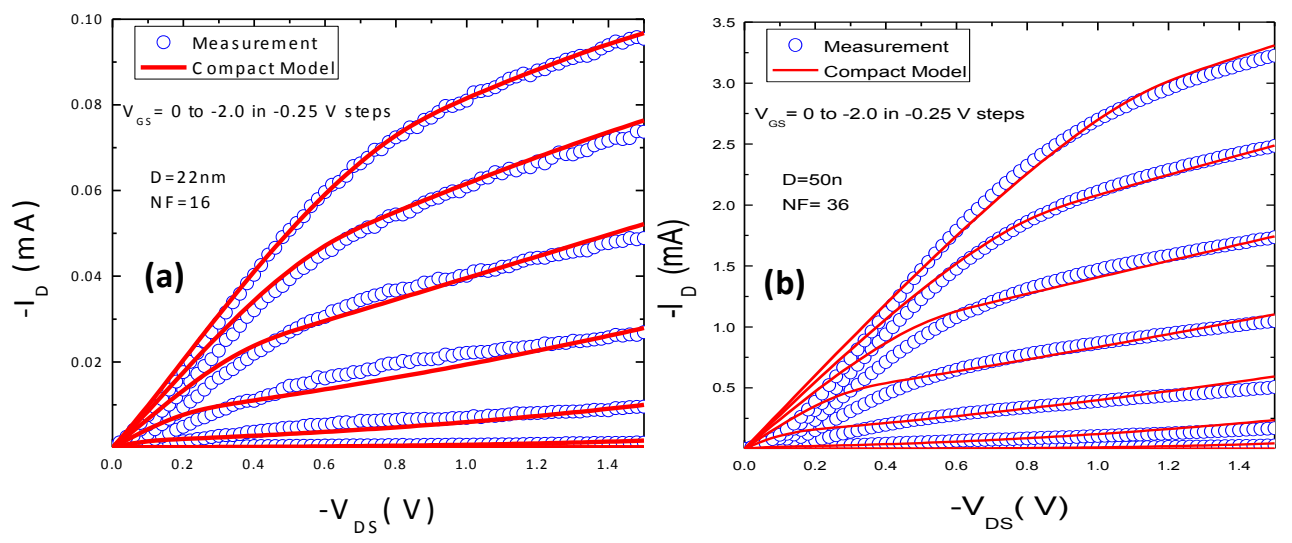

Fig. 3: $I_D$-$V_{DS}$ of JLNTs with (a) 22nm diameter and 16 nanowires in parallel and (b) 50nm diameter and 36 nanowires in parallel.

## III. LOGIC PERFORMANCE ASSESSMENT

In this section, we leverage the developed compact model to assess the performance metrics of various topologies of an elementary logic gate in the VNWFET technology. While it is possible to simulate logic gates implementing multiple Boolean operations, we focus in this paper on the comparison between several topologies implementing a single inverter operation. This is partly due to the lack of experimental devices and consequently measurements with which the compact model parameters can be defined; but it also targets a full understanding of the relative merits and drawbacks of the device itself, minimizing design-specific issues. In a first exploration we assess the simulated performance of p-type only inverters, while in a second exploration, using a literature survey, we extrapolate the model to n-type VNWFETs in order to explore a simple complementary inverter structure. Finally, we establish a comparison with the 7nm FinFET technology node using typical values, in preparation for further analysis in section IV.

### A. *Simulated structures*

This work focuses on inverter structures implemented with passive- and active-load topologies, as well as with complementary topologies.

#### 1) *P-type only inverters*

P-type only structures use a p-type device as conventional pull-up, and implement the pull-down branch as a resistance, with either a passive (fig. 4(a)) or an active (fig. 4(b)) load. For the latter, we use a p-type device configured as current source. In both cases, the pull-down load is responsible for the low logic state output. This type of structure is known to be less efficient than their complementary counterparts, but firstly enables validation of the use of experimental data for designing logic, and secondly gives first insights into the use of such devices.

#### 2) *Complementary inverter*

For the complementary circuit, based on [9][10] and shown in fig. 4(c), we conjecture a value for the carrier mobility in the n-type VNWFET channel and consequently its drive current. This value is 3x that of the p-type VNWFET. Hence to balance the circuit for a switching input voltage value halfway between the supply rails and for roughly equivalent noise margins, we target identical currents in both devices. To achieve this, we set the NF (number of nanowires) per device in the P-type equal to 3x that of the n-type.

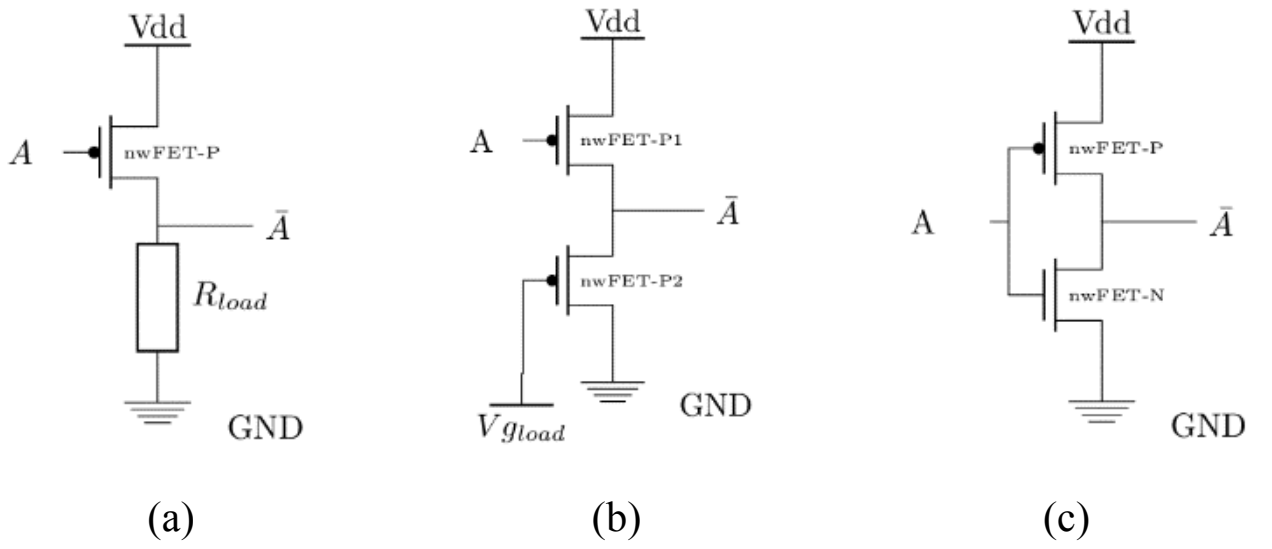

Fig.4: Schematics of the studied inverter structures, (a) passive load, (b) active load and (c) complementary

### B. *Results*

The goal of the following simulation-based exploration is to study the impact of using a large range of nanowires per transistor on typical static and dynamic logic performance metrics. In the simulation protocol, we assume that the gate capacitance behaves in the same way for both p- and n-type VNWFETs, and that the capacitive load on the output of each structure is equivalent to its own theoretical input capacitance (i.e. fanout of 1). Since the p-type VNWFET gate capacitance with NF=16 is experimentally determined to be 50aF, and assuming that its evolution with NF is linear, we deduce a capacitance contribution per nanowire of 3.25aF. Measurements were performed using the model described in section II as implemented in Verilog-A and simulated using the Spectre™ commercial simulator.

#### 1) *Static performance*

DC simulation points enable the extraction of typical static characteristics of the p-type VNWFET transistor.

##### a) *$I_{on}/I_{off}$ ratio*

In this analysis, we characterize the p-type VNWFET characteristics in terms of $I_{on}/I_{off}$ ratio for values of NF ranging from 3-300. To measure $I_{on}$ (resp. $I_{off}$), we set input A=0 (resp. A=1) such that the pull-up device is on (resp. off) in all structures.

We observe a linear increase in the leakage current with NF at a rate of 61pA per nanowire. Given the 16nm nanowire diameter, this translates to 0.3µA/µm$^2$ leakage current density in the p-type VNWFET. However, device $I_{on}$ does not increase linearly with NF – in fact, the rate of increase slows down when using large values of NF. As a result, device $I_{on}/I_{off}$ ratio *decreases* with increasing NF, from $15\times10^3|_{NF=10}$ to $6.5\times10^3|_{NF=300}$.

##### b) *Logic level degradation*

The load in the pull-down branch of both passive- and active-load inverters is a major factor both for logic '1' level degradation and for high-low propagation delay. Its value is a tradeoff: increasing the load decreases logic level degradation but increases propagation delay. For the studied structures the best compromise, as shown in fig. 5, gives a 15% logic '1' level degradation and a 15% logic '0' level overshoot during high-low transitions at the output for a 1GHz input signal.

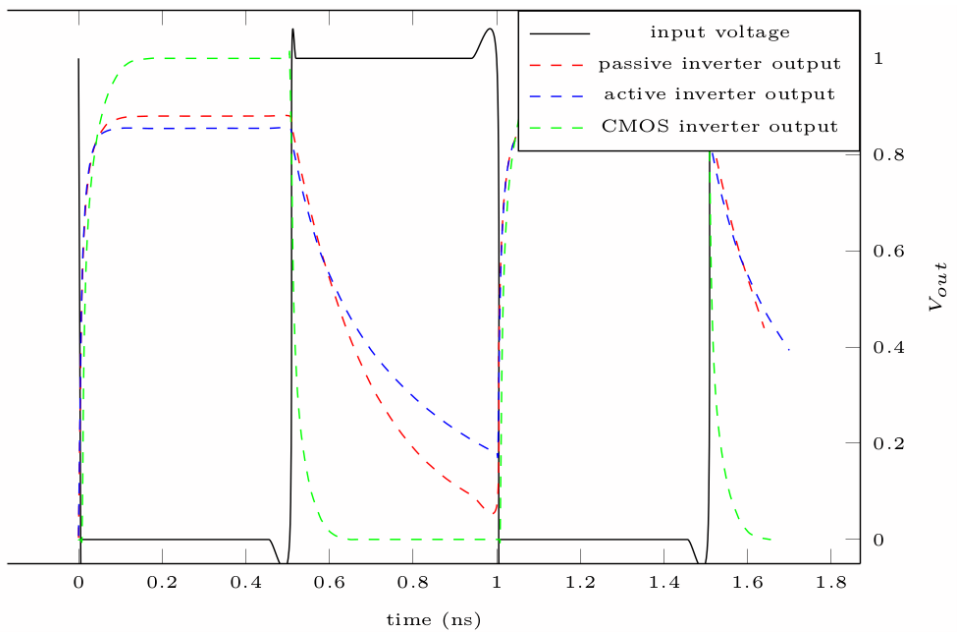

Fig. 5: Voltage across the load capacitor for the studied structures

#### 2) Dynamic performance

As shown in fig. 5, we carried out transient simulations to extract relevant dynamic performance metrics, using a 1GHz data input with rise and fall times equal to 10ps. As previously indicated, each inverter shown in fig. 4 was simulated with a fan-out of 1.

##### a) Propagation delay

For small values of NF, we observe a delay (measured as $t|_{Vout=50\%Vdd} - t|_{Vin=50\%Vdd}$) ranging from 5-10ps according to the type of inverter (the lowest delay is achieved by the passive load inverter). When increasing NF, the gate delay increases. This result can be linked principally to the sublinear increase in $I_{on}$ with NF, and the linear increase in gate capacitance with NF.

##### b) Dynamic energy consumption

We also measure the energy required to transit through the transistor channel when changing state. We calculate the amount of charge for a low-high transition at the output for the self-loaded complementary inverter. This value varies linearly with NF and works out to 11aC per nanowire. With a 1V supply voltage this gives us an energy consumption of 11aJ per nanowire for a low-high transition at the output.

#### 3) Fanout analysis

Due to the sublinear variation of $I_{on}$ with NF, self-loaded logic cells with high values of NF cannot charge completely in the available time (fig. 6). At 1GHz and for NF>300, an inverter cannot cascade with an identical logic cell. Similarly, when increasing the fan-out (number of cells controlled by the inverter), this boundary reduces until fan-out = 5, where a single nanowire transistor cannot drive 5 identical cells simultaneously. This information is crucial regarding power and delay management when designing larger cells.

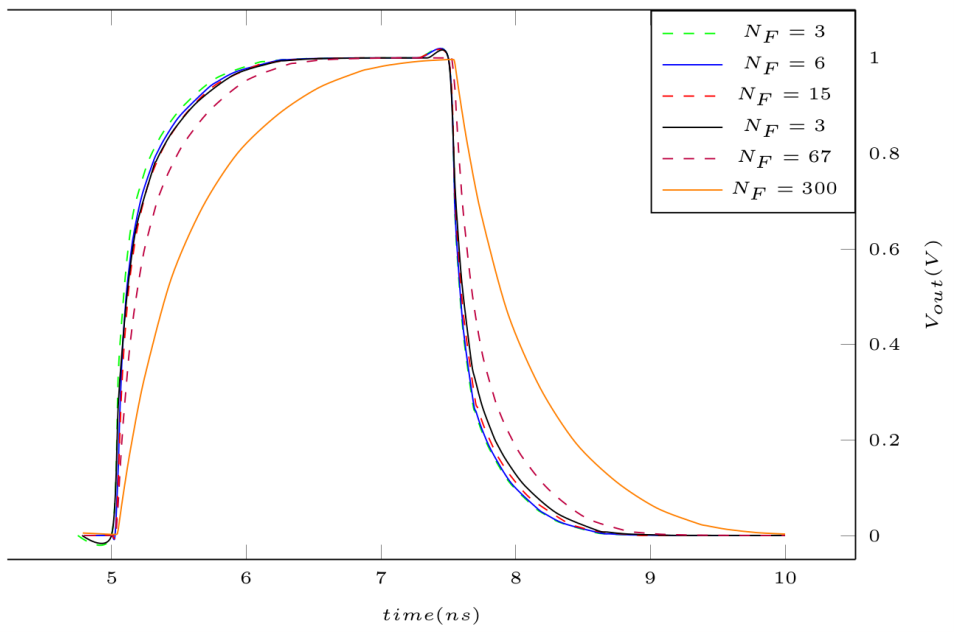

Fig. 6: Voltage across the load capacitance of the self-loaded complementary inverter according to NF (p-type VNWFET)

### C. Comparison with FinFET and conclusion

Based on the previous results, there is a clear advantage for using logic cells with low NF, both for power consumption concerns and for fan-out. The values obtained are compared in Table I to FinFET values from the literature, both for static values [11] and for propagation delay [12].

TABLE I. VNWFET / FINFET COMPARISON

| Metric | VNWFET | 7nm FinFET |
|---|---|---|
| Static leakage current density (µA/µm$^2$) | 0.3 | 1 |
| $I_{on}/I_{off}$ ratio (*10$^4$) | [1.5 – 0.65] | ~8 |
| Propagation delay (ps) | [5 - 10] | 2.2 |

## IV. 3D LOGIC CELLS

### A. Going vertical: Implications on physical circuit design

#### 1) Paradigm change

Vertical transistor channels lead to a paradigm change in the design of logic cells. Source and drain contacts, separated by the vertical channel, can occupy the same lateral space. Stacked series transistors further improve the gain in circuit density. Further, the additional dimension enables numerous spatial configurations for the same logic functionality [13]. However, careful evaluation of gate contacts and routing is necessary to ensure the best tradeoff between density and performance. In this section, we identify critical dimensional constraints, formulate λ-rules for the VNWFET technology and leverage them to compare footprint to lateral FinFET technology.

Although this article does not aim to explore complex logic structures using this technology, it lays the foundations for carrying out a complete and exhaustive study with this objective. For this reason, in order to deal with the significant differences between planar FinFET technology and vertical nanowire technology, initial designs must share as much common ground as possible with a tried-and-tested yet cutting-edge technology. Based on the comparison results we extrapolate the comparison metrics to projected figures considering using the potential of the VNWFET technology to its full extent.

#### 2) Comparison basis

A planar FinFET channel is composed of a number of fins, according to the desired transistor characteristics (fig. 7). Similarly, a VNWFET channel is composed of several vertical nanowires. In this work, we aim to compare the footprint of VNWFET-based logic cells with respect to FinFET-based logic cells. We take as baseline reference λ-rules for elementary standard cells based on the 7nm FinFET technology [14] established in the context of exhaustive layout and performance benchmarking. Lambda-based rules (λ-rules) constitute a simple tool that allows first order scaling by linearizing the resolution of the complete wafer implementation. While modern processes rarely shrink uniformly, λ-rules remain useful to make first-order cross-technology spatial comparisons.

The principle of λ-rules is to decorrelate characteristic sizes from absolute dimensions by expressing them as a function of some reference length unit (λ). The λ value used for the FinFET represents twice the fin thickness ($T_{si}$ – as shown in fig. 7), which represents the smallest mask dimension (oxide thickness, established through epitaxial growth, is not correlated to lithography or mask limitations). Correspondingly, the smallest dimension in the VNWFET transistor is the nanowire diameter D (see fig. 8) and is accordingly used to define λ for the VNWFET technology. An important observation in both technologies is that dimensions in the transistor zone are comparable to λ, while dimensions comparable to 3λ are used in the routing and contacting of the transistor.

In the baseline reference, FinFET planar transistors are at the 7nm node, such that λ=3.5nm, while the current state of VNWFET technology allows a minimal nanowire diameter such that λ=16nm. It should be stressed that this is representative of an emerging research technology under development rather than an inherent limitation to the technology.

Table II shows the λ-rules as established in [14] for 7nm FinFETs as well as those chosen in this paper for VNWFETs.

TABLE II. LAMBDA-RULE COMPARISON BETWEEN FINFET AND VNWFET

| Parameter | Value in 7nm FinFET (nm) | Value in projected VNWFET (nm) | Comment |
|---|---|---|---|
| $T_{fin}$ / $D$ | $3.5 = \lambda_{fin}$ | $11 = \lambda_{NW}$ | Fin thickness / nanowire diameter |
| $T_{si}$ | $2{*}\lambda_{fin}$ | | Fin length |
| $H_{fin}$ / $H_{NW}$ | $4{*}\lambda_{fin}$ | 30 | Height |
| $T_{ox}$ | 1.55 | 5 | Oxide thickness |
| $P_{fin}$ / $P_{NW}$ | $2{*}\lambda_{fin} + T_{fin}$ | $2{*}\lambda_{NW} + D$ | Pitch |
| $W_C$ | $3{*}\lambda_{fin}$ | $3{*}\lambda_{NW}$ | Contact size |
| $W_{M2M}$ | $2{*}\lambda_{fin}$ | $2{*}\lambda_{NW}$ | Gate to contact space |

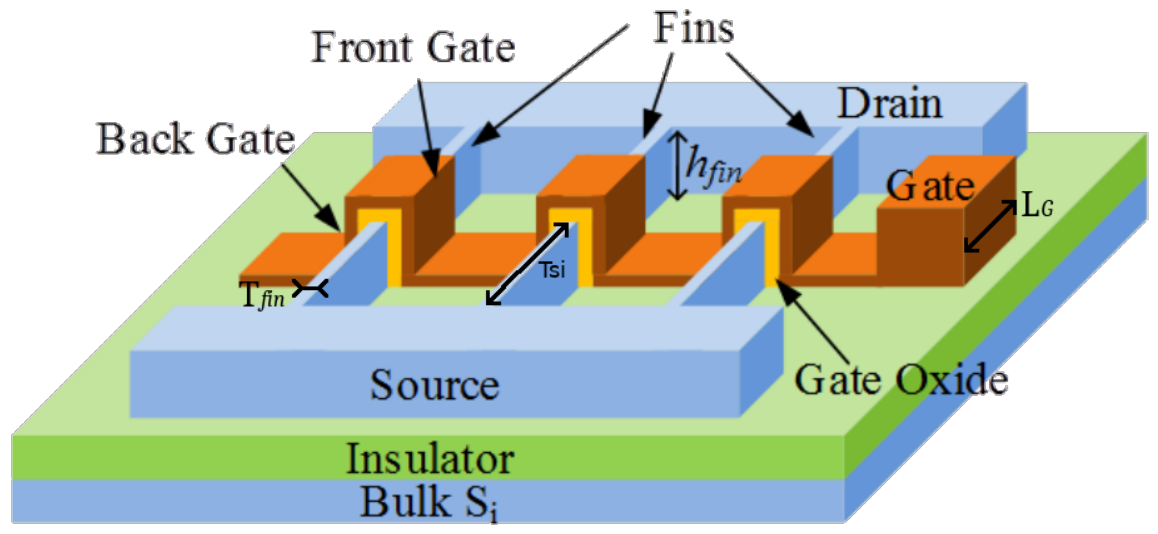

Fig. 7: Perspective view of a 7nm node FinFET transistor [14]

## B. VNWFET technology and device structure

There are 3 metallic contacts along the transistor’s vertical channel. This structure is shown in fig. 8 [2]:

- The bottom PtSi contact surrounds the bottom of the nanowire and establishes a first access to the transistor channel. This contact is ultimately used as drain or source.
- A top Al contact covers the top of the nanowire and establishes a second access to the transistor channel. This contact is similarly used as a drain or source.
- A Cr layer in the middle surrounds the center of the nanowires and is separated from the silicon by a gate oxide. This metal contact acts as the gate.

It is worth noticing that the gate structure surrounds the channel, thus categorizing this type of transistor as a Gate-all-around (GAA) FET. Moreover, as compared to FinFET technologies, the silicon in the region of the drain and source is doped in the same way as for the channel zone. Specifically, the nanowire to which the drain, gate and source are attached is etched in a uniformly doped silicon bulk [1]. This transistor is thus also a junction-less transistor.

The perspective view of the nanowire transistor provides insights into the tridimensional structure of the device. Since we focus on the lateral footprint and for the sake of improved visibility, vertical dimensions are not to scale in this view. A single transistor may comprise multiple (NF) nanowires in its channel and each nanowire is surrounded by gate oxide before any contacts other than the bottom contact is deposited. Spacers made of oxide are represented between top contact and gate and between gate and bottom contact to isolate those metals.

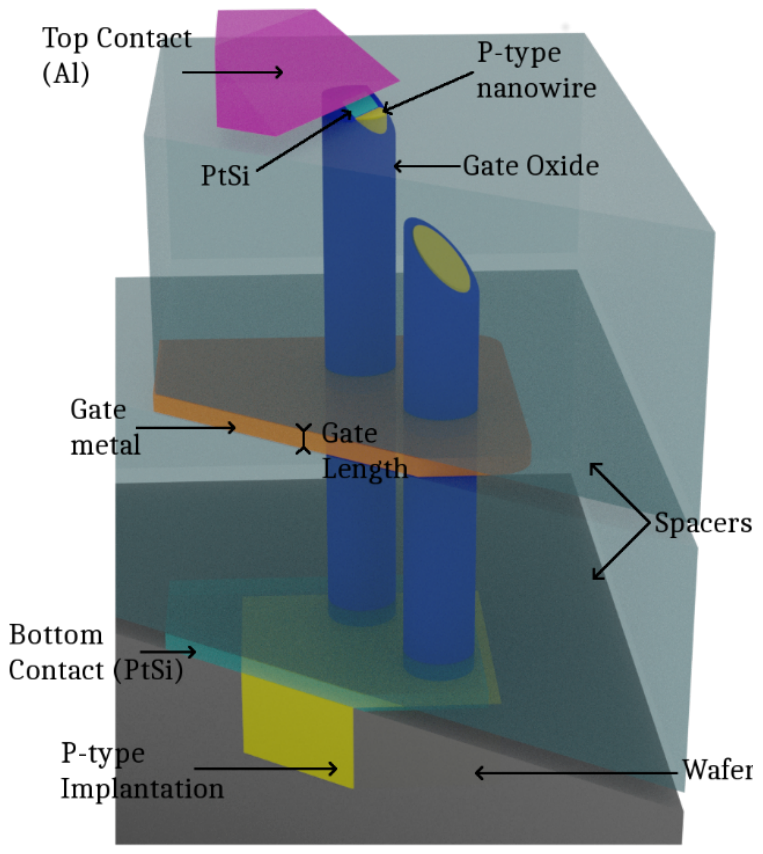

Fig. 8: Perspective and cut view of a projected VNWFET transistor

In order to facilitate the differentiation between p- and n-type, and for the sake of clarity, the figures representing VNWFET structures in the remainder of this paper will show neither the gate oxide surrounding the silicon nanowire nor the insulating spacers for each metallic layer.

## C. Footprint estimation

As indicated previously, we focus on the footprint (lateral area) in order to keep common grounds with the FinFET technology. The vertical height of the logic cell is considered unimportant in this comparison and unrelated to any FinFET dimension. The same is true for gate length and contact thickness.

In the standard cell approach, the lateral “height” of the cell (i.e. the distance between supply voltage and ground) is constant for all standard cells in a technology. Data inputs and outputs are typically located in the middle of the cell. Their position is not constrained and their access is not taken into account in the cell design. This type of layout allows the designer to assemble each logic gate on the same level with only inputs and outputs to route properly, usually through a dedicated routing channel.

Several important points mentioned in section III are taken into account in order to implement logic cells in the context of standard cell design. The current technology is used for characterization and trials on vertical nanowires. It is thus unable to sustain the requirements of the λ-rule constraints we introduced in Table 2. Indeed, the 16nm diameter value is not the main concern, since the dimensions of the metallic layers are much larger for electrical characterization purposes and manufacturing process limitations. The lithography equipment used in the process is not intended for this scale of precision.

These manufacturing changes, as compared to [1], remain credible in a foreseeable future. In a nutshell, the main assumption is that contact and gate dimensions are in the same range as the nanowire dimensions.

## D. Layout footprint comparison example

The comparison method explained above is applied to a CMOS inverter structure with balanced switching corresponding to nanowire mobility [9].

The difference between the mobility of both transistor channels suggests that the n-type and p-type transistors respectively possess one fin or one nanowire and one fin or three nanowires. In fig. 9, the layout is composed of large voltage supply extensions for both $V_{DD}$ and GND and an active zone where the transistors are connected together. The layout footprint is the product of the width and length given in λ values.

In both Figs. 9(a) and (b), we notice a similar inverter structure, where the position of the supply voltages and input/output are indicated.

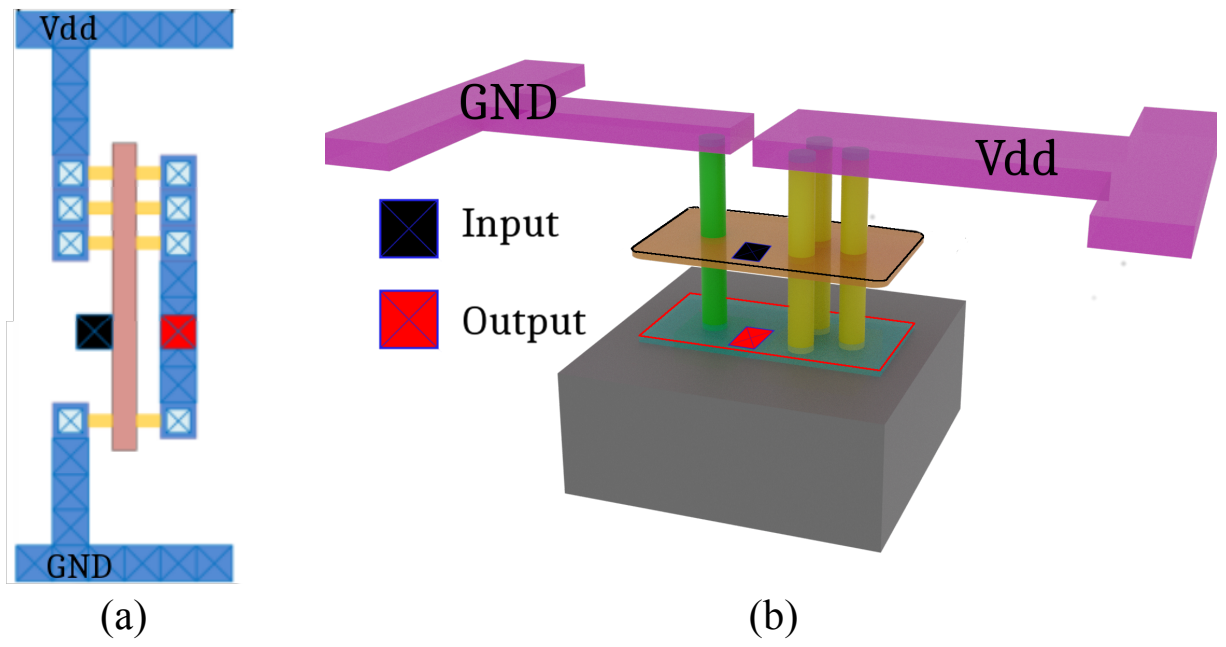

Fig. 9: (a) Layout (top view) of a 7nm FinFET inverter gate [Cui14], (b) Perspective view of a VNWFET inverter gate according the projected technology

The FinFET inverter footprint is 48λ high and 18λ wide while the VNWFET inverter footprint is 31λ high and 15λ wide. This represents a 48% footprint area reduction. If we choose a less restrictive comparison criterion and consider the active part alone (removing the 12λ supply contacts for both layouts), we observe an 84% footprint area reduction.

## V. Insights for Large Scale Integration

While devices based on vertical nanowires have been compared to lateral GAA devices in the past [15], this work used apparent mobility differences between both device channels to justify the difference in gate lengths. In order to achieve similar drive strength for both devices, the vertical device gate length is around 2× that of the lateral device. Such electrical considerations help to set the vertical dimensions for the VNWFET. Gate length values in the referenced article are in the 10-20nm range. This value fits our designs without any impact on the device footprint area and its impact on overall performance will be studied in the future. The fact that VNWFET dimensions such as gate, spacer and channel lengths are decorrelated from the lateral footprint allows electrical parameters to be tuned without touching the cell design. This favors the standardization of simple cells, as well as the achievement of complex logic design and scalable electronics. In this work, the comparison method separated electrical behavior concerns from device layout to establish a workflow and to enable the future consideration of stacked-gate vertical devices [16]. Stacking gates requires device $I_{on}$ to be high enough to drive the whole common channel. We demonstrate in section III that with current technology, the fan-out limit is 4 for a very low number of transistors. Thus, we can expect at least a functional 4-stacked gates transistor, which already enables significant opportunity for disruptive logic designs.

## VI. Conclusion

This work considers the use of VNWFETs as a means to implement 3D logic blocks. We have built a technology scalable physics-based compact model and implemented it in Verilog-A as incorporated in a dedicated circuit design workspace. This environment has been used to simulate innovative 3D layouts of inverter cells. The layout of the complementary inverter has been compared with projected 7nm FinFET technology through the use of λ-rules. We showed that the VNWFET-based approach achieves 48% footprint reduction and can reach 84% if only the active part is considered. Beyond λ-rule comparisons, we presented another physical layout implementation that leverages the unique features of VNWFETs, where dimensions such as gate, spacer and channel lengths are de-correlated from the transistor footprint. This important property allows electrical parameters to be tuned without any impact on cell design. The standardization of such simple logic cells will pave the way for more complex VNWFET logic cell designs.

## Acknowledgments

This work was supported by the French RENATECH network (French national nanofabrication platform) and by the LEGO project through ANR funding (Grant ANR-18-CE24-0005-01).